# A Taxonomy of Response Strategies to Toxic Online Content: Evaluating the Evidence

***Lisa Schirch*, Kristina Radivojevic*, Cathy Buerger*****
***University of Notre Dame***
****Dangerous Speech Project***

## Abstract

Toxic Online Content (TOC) includes messages on digital platforms that are harmful, hostile, or damaging to constructive public discourse. Individuals, organizations, and LLMs respond to TOC through "counterspeech" or "counternarrative" initiatives. There is wide variations in their goals, terminology, response strategies or methods of evaluating impact. This paper identifies a taxonomy of online response strategies, which we call *Online Discourse Engagement (ODE),* to include any type of online speech to build healthier online public discourse. The literature on ODE makes contradictory assumptions about ODE goals and rarely distinguishes between them or rigorously evaluates their effectiveness. This paper categorizes 25 distinct ODE strategies, from humor and distraction to empathy, solidarity, and fact-based rebuttals, and groups these into a taxonomy of five response categories: defusing and distracting, engaging the speaker's perspective, identifying shared values, upstanding for victims, and information and fact-building. The paper then systematically reviews the evidence base for each of these categories. By clarifying definitions, cataloging response strategies, and providing a meta-analysis of research papers on these strategies, this article aims to bring coherence to the study of ODE and to strengthen evidence-informed approaches for fostering constructive ODE.

## Introduction

Despite disagreement on its causes or scope, there is wide public concern about toxic online content (TOC) and growing interest in prosocial online discourse engagement (ODE). There is no consensus on the definition of TOC or response strategies for ODE. People infer different meanings to terms like "counterspeech" and "counternarrative." While all may define this type of ODE as simply "a response to harmful comments", there are many different ways to respond. In the course of our research, individuals and groups who respond to TOC hold different theories of change, or assumptions about what is effective (Seehausen et al., 2012; Hangartener, et al., 2021; Buerger, 2022; VODA, 2025). Most research teams designing LLM chatbots to support ODE rely on a "single shot" or "few shot" approaches offering  a simple prompt to "respond" to an example of TOC or a few examples of responses to TOC. As more individuals, organizations, and automated systems take part in ODE, there is a need for a comprehensive taxonomy of response strategies and an assessment of the evidence of their impact.

***In this work, we define TOC as an umbrella term that includes, but is not limited to, intolerance, hate speech, fear speech, violent threats, toxic polarization, and divisive information.*** The European Court of Human Rights and the United Nations define *hate speech* as content that spreads hatred based on intolerance, discrimination, attacks, or inciting violence against groups based on religion, race, ethnicity, nationality, sexuality, disability, or class (ECHR

2023). *Fear speech* exaggerates threats from a community as a perpetrator, aiming to influence beliefs about crime, immigration, or political opponents (Saha et al., 2023). Toxic polarization goes beyond simply holding different views toward an us vs them approach that dehumanizes or demonizes "them." *Divisive information* is politically contested informatio*n. Disinformation* is often deployed as part of a political, economic, or ideological strategy (Wardle, 2024).

As the scope and scale of harmful online content grew, some groups began using the concept of "toxic speech" as an umbrella term. Google's Perspective API social media toxicity detector, defining toxicity as "A rude, disrespectful, or unreasonable comment that is likely to make people leave a discussion" (Jigsaw, 2022). The Civic Health Project defined *toxic speech* as speech that encourages partisan violence, spreads biased views of opposing groups, expresses moral outrage that drives polarization, or uses dehumanizing language (Civic Health Project, 2025).

Online content moderation led to broad concerns about free speech. Legal scholars cite a long history of the "counterspeech doctrine" dating to Supreme Court Justice Louis Brandeis's 1927 opinion that censorship of harmful speech should only happen if and when the harms could not be addressed through *additional speech*. A law review article entitled "Counterspeech 2000" summarized this as the solution to *bad speech* is *more speech* or "counterspeech" in a so-called "marketplace of ideas" (Richards & Calvert, 2000). While social media platforms developed community guidelines to define unacceptable online content, enforcement is weak (CCHD, 2021). Research on interventions to *nudge* users to follow these guidelines, such as popup reminders of the guidelines, have demonstrated positive behavioral impact (Matias, 2019).

This paper offers a conceptual framework of the diverse approaches to online public discourse-based response strategies to TOC both offline and online. *Counterspeech* refers to reactive, immediate, direct responses to hateful, harmful, or toxic speech (Bartlett & Krasodomski-Jones, 2015). *Counternarratives* are broader responses to a set of ideas (Dangerous Speech Project 2025). Rather than a brief or reactive reply, they provide alternative stories, frames, or identities often centered on tolerance, coexistence, and dignity, that challenge the worldview or ideology embedded in toxic content (Belanger et al., 2023). Others refer to ODE as "online civic intervention" (Porten-Cheé et al., 2020), positive speech or "alternative narratives"(Choquette et al., 2024) or "positive narrative expansion"(Kruglova & White, 2025). In this article, we use the term *Online Discourse Engagement (ODE)* as a broader umbrella term that includes both counterspeech and counternarratives, but also extends beyond them to include any type of online speech to build healthier online public discourse.

Figure 1 illustrates that ODE is a general response to TOC in online social media comment sections. In some cases, ODE can also be used more generally to support public discourse norms without being in response to a specific piece of content or narrative. For example, ODE migh take place after an incident of public violence without being a direct response to TOC. This paper uses both acronyms TOC and ODE as short-form.

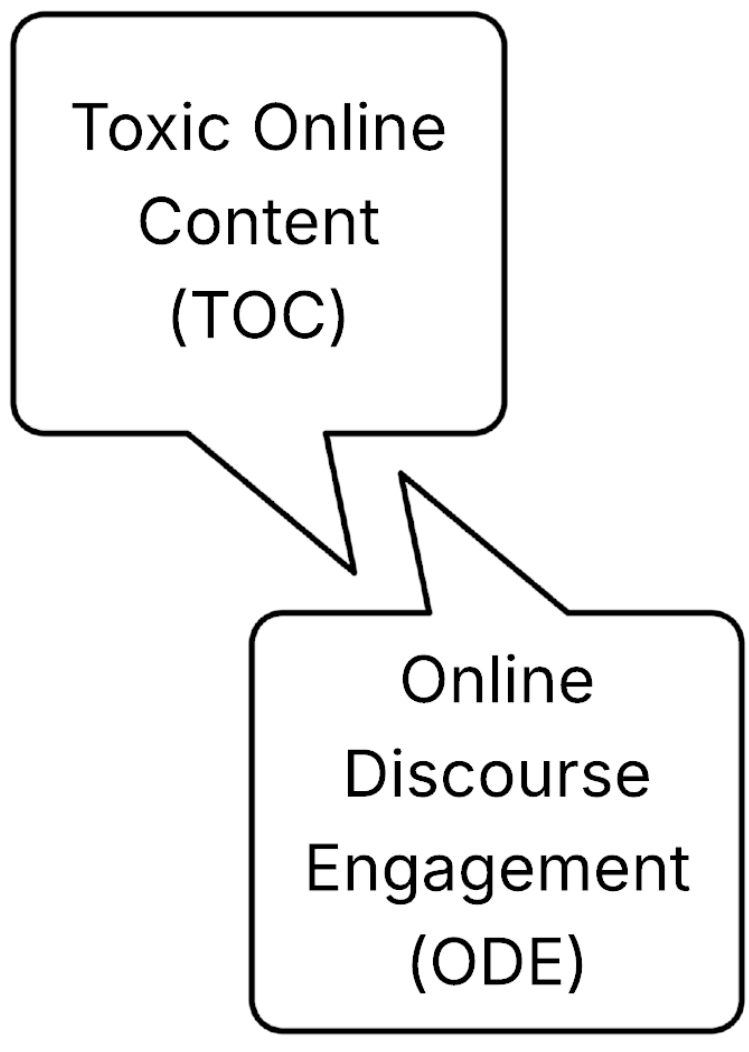

**Figure 1:** Visual representation of the conceptual framework used in this paper.

## Research Methodology

In this research project, we employed a mixed-methods approach that combined action research, interviews, focus groups and literature reviews to answer two research questions:

***(RQ1) What is a comprehensive taxonomy of online discourse engagement (ODE)?***

***(RQ2) What is the evidence base to evaluate the efficacy of each ODE strategy?***

First, our team drew on action research based on facilitating over 40 meetings or dialogues between groups in conflict. This action research methodology draws on decades of research on prosocial public discourse drawing from our own research team's expertise in conflict resolution, dialogue, and negotiation communication strategies. These include skills like asking clarifying questions, active listening through paraphrasing and reframing, finding shared values or common ground, and modeling care and empathy for others (Rosenberg, 2003; Search for Common Ground, 2021). We also drew from the fields of preventing and responding to violent extremism, with its emphasis on counternarratives that address core grievances and urge help-seeking behaviors (Schirch et al., 2023). Initial mapping of ODE response strategies to TOC fell into three broad categories relating to responses directed toward the victims, bystanders, or perpetrators of toxic speech online. An early taxonomy identified thirteen strategies, such as empathy, solidarity, offering facts, defusing, and paraphrasing.

Next, we conducted interviews and focus groups with key individuals and groups involved in counterspeech from January to March of 2024. These groups included #Iamhere, Dangerous Speech, Search for Common Ground, Build Up's project on The Commons, Smart Politics, and the Civic Health Project in addition to individual experts from the field of conflict resolution. Based on this research, we added additional strategies to our initial taxonomy.

Next, we expanded our taxonomy further with the Dangerous Speech's taxonomy (Benesch 2016, Buerger 2024) and a literature review of counterspeech strategies (Fraser et al., 2023) including a 2023 study of responses to TOC on X (Yu et al., 2023). The research team added to and refined a total of 25 response strategies divided into five categories.

In the final stage of our research, we used our taxonomy as search prompts for keyword searches on Google Scholar, ArXiv, and Google Search. Graduate students categorized articles, building a database of peer-reviewed articles on ODE strategies. Students followed a rulebook that directed them to begin their search from fields such as counterspeech, peacebuilding, communication studies, social psychology, and digital governance. Students coded the research based on levels of evidence, and number of citations.

## Conceptual Challenges

As with any emerging field of study, there is little clarity or standardization in terminology, conceptual frameworks, taxonomies, and metrics among those who practice or study ODE (Chung et al., 2024). Scholars, practitioners, and funders of ODE face several conceptual challenges.

First, there is little agreement on terminology, with overlapping terms such as counterspeech, counternarrative, online civil intervention, civility and other terms all being used to describe prosocial responses to toxic online content. Second, practitioners and researchers are often vague about what strategies are included. While there are a few taxonomies of approaches used in the wild, there has been no thorough collection of the wide variety of strategies employed or what evidence exists of their impacts.

Practitioners use strategies that they believe are effective. Others learn about ODE through guidebooks by organizations that compile practitioner strategies and informal lessons learned. Researchers study some of the strategies, but not others. Research studies often focus on a limited set of ODE strategies, such as fact-checking, confronting perpetrators, or expressing solidarity with victims. Some research papers evaluating ODE do not even define the response strategies they are measuring, assuming that “counterspeech” is self-evident. Many authors define these terms by *what* they do (responding to TOC) rather than *how* they do it (response strategies). Some academic papers study “empathy” as a strategy, but it is unclear whether they are referring to empathy toward the victims of toxic speech or empathy toward perpetrators of toxic speech.

Second, ODE practitioners hold different goals, but often these goals are not explicit. The overall goal of most counterspeech efforts is to promote *healthy public discourse norms* to foster respectful, inclusive conversations online that encourage deliberation and diverse viewpoints. *Public discourse norms* are the informal rules that guide how people speak and listen in civic spaces, emphasizing respect, fairness, and constructive engagement. These norms directly connect to the concept of the *prosocial*, which refers to behaviors and designs that support cooperation, empathy, mutual understanding, and the well-being of the community and social cohesion (Schirch, 2025).

ODE can also include promoting adherence and accountability to community standards and rules on some platforms. These rules usually define the edge of acceptable behavior and aim to reduce hateful, harmful, and toxic content. Prosocial ODE approaches aim not only to reduce harm but also to foster mutual respect and shared values.

Individuals or groups involved in ODE may have diverse ways of achieving prosocial ODE. For some groups, ODE goals aim to change the minds and behavior of those who post TOC. For others, they look to reducing the negative impacts on TOC victims (Buerger 2021). Some ODE initiatives aim to influence byscrollers (online bystanders), while others focus on victims, and some on perpetrators, as can be seen in Figure 2.

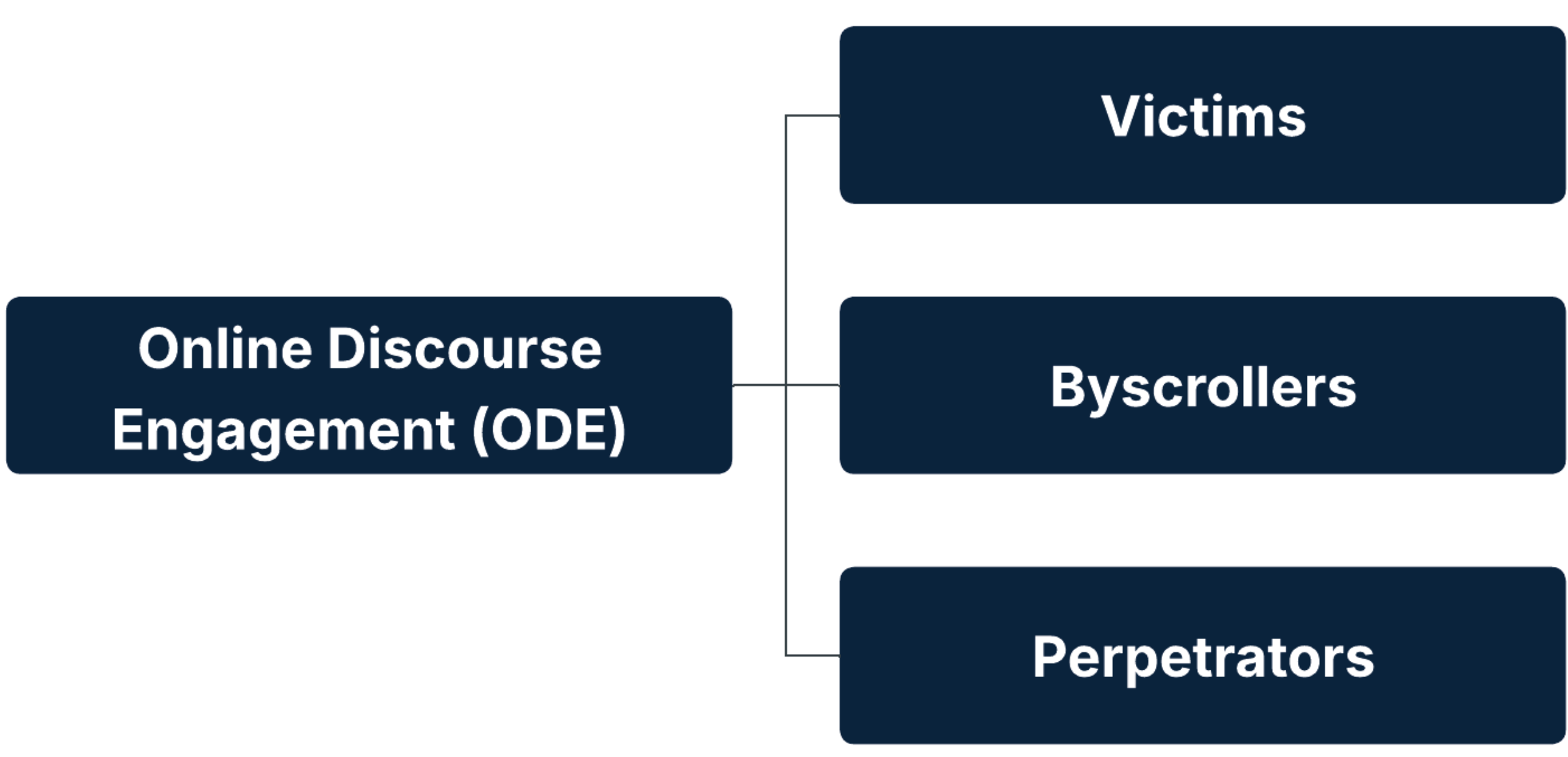

**Figure 2:** Visual representation of different intended audiences of ODE.

Some ODE groups such as *Civic Health Project* responds to individual perpetrators who seek fame, attention, or profit by calling out bot-like or fraud-like behavior. Others such as *Institute for Strategic Dialogue* focus on organizing counternarratives to collective or coordinated extremist groups, political parties, or foreign states that promote TOC as a means of recruiting new members, shifting public opinion, mobilizing violence against minority groups, or dividing a society. Even when individuals or groups engage in ODE with the same general audience in mind, they may have different goals, as can be seen in Table 1.

**Table 1:** Different types of ODE goals, as linked with the intended audience.

| Victim-Oriented Goals | Byscroller-Oriented Goals | Perpetrator-Oriented Goals |
|---|---|---|
| Protect and express empathy and solidarity with this group | Dissuade byscrollers from using toxic speech | Challenge, correct, and deter toxic speech |
| Amplify positive narratives about the group | Equip byscrollers with critical thinking skills to identify and expose toxic online content, including manipulation, propaganda, or misleading content | Persuade and encourage more productive communication |
| | Encourage byscroller intervention rather than silence or acceptance by modeling ODE | Deter and interrupt the process of online recruitment and radicalization by undermining the appeal and credibility of extremist ideologies. |

A third research challenge is that practitioners and scholars measure impact in different ways. Some ODE initiatives measure effectiveness based on social media "impressions" or how many people viewed the comment. In a review of research on counternarratives for preventing and countering violent extremism, researchers found a reliance on engagement metrics such as impressions, noting this has little value in indicating if real attitude, belief, or behavior change has occurred (Helmus & Klein, 2018). While major social platforms measure impressions and engagement (likes, shares, or more comments), this type of measurement does not indicate what might be causing impressions or engagement. A post that makes people laugh or angry might get many impressions, but not have any significant effect in promoting prosocial public discourse norms. Yet measuring changes in the attitudes, beliefs, and behaviors of perpetrators of TOC, victims of TOC, or byscrollers is difficult and has rarely been carried out. Ideally, researchers would compare different strategies to learn if a perpetrator takes down a comment with TOC, edits the comment, and/or does not post further TOC. Such research could also measure the toxicity of a comment thread after one or more ODE comments.

Research might also rely on interviews or public surveys to find out if victims of TOC believe the ODE is important or effective in reducing harm to them. One study finds that different identity groups seem to prefer different ODE strategies (Matthew et al., 2019). And a similar qualitative methodology could help determine what, if any, impact different ODE strategies have on byscrollers. The lack of a comprehensive ODE taxonomy weakens research on and the practice of ODE.

## A Taxonomy of Online Discourse Engagement (ODE) Strategies

Based on our interviews, focus groups, and an initial literature review, we developed a taxonomy of five main categories and 25 distinct ODE response strategies that fall within them. Table 2 illustrates these categories and strategies. Whereas earlier taxonomies had included memes and GIFs as separate strategies, we viewed these as methods of delivery rather than separate strategies. Humor, for example, is a strategy communicated either via text or visual methods. Such a taxonomy is important for several reasons. Without a more robust taxonomy, researchers often fail to distinguish what exact strategy they are studying. This creates confusion, when some researchers assume counterspeech is about empathy to victims while others assume it means empathy to perpetrators, and still others assume it means facechecking.

Without a taxonomy, researchers cannot test different response strategies to determine which are most effective. Without knowing more about what strategies are effective, individuals volunteering to participate in counterspeech operate with little guidance. While groups like #iamhere have thousands of volunteers, only a small minority participate each week (Buerger 2021); possibly because they require more capacity building and support. Without evidence of which strategies are most effective, volunteers may inadvertently rely on strategies that have no effect or even a negative effect. A number of studies found that there are significant problems with counterspeech actually amplifying toxicity rather than decreasing it. In one study of Reconquista Internet (RI), the actual practice of responding online mirrored rather than challenged negative tones. These researchers assert that counterspeech did not reflect constructive communication or improved understanding but instead often relied on impulsive, unconsidered reactions (Keller & Askanius, 2020). Other studies of the same group RI found that the group's efforts were effective, as discussed later in the paper (Garland et al., 2022). This points to the next challenge. Without the comparative research that a taxonomy allows, AI researchers attempting to build automated ODE into social media platforms also operate blindly, without knowing how best to train their LLMs to respond to TOC.

**Table 2:** Taxonomy of five main categories and 25 distinct ODE response strategies that fall within them.

| Defuse and Distract | Engage Speaker's Perspective | Identify Shared Values | Upstand for Victims | Information and Fact Building |
| --- | --- | --- | --- | --- |
| Use humor | Reframe to express grievance in a less harmful way | Establish safety by modeling | Show empathy or solidarity to victims | Offer news links that show different |

|  |  | constructive discourse |  | perspectives; diversity |
|---|---|---|---|---|
| Mock or ridicule | Show comprehension with paraphrasing | Set dignity boundaries | Invite others to upstand | Offer science or legal evidence |
| Expose profit or bot behavior | Show care and empathy to the speaker | Identify shared values or ideas | Denounce and expose toxicity | Erode the speaker's intellectual framework; point out hypocrisy |
| Encourage help-seeking | Ask questions, perhaps to provoke reflection | Use moral reframing | Warn of consequences | Show inconsistencies in logic on and/or ask questions to prompt uncertainty about speaker's claims |
| Redirect with breathing, resilience messaging or other calming content | Show support for certain parts of a post | Use religious values | Chime in on another's norm-setting | Prebunking or Debunking Universal Traits: Stating that the stereotypical trait of a victim group is not unique to the target group |
| *Memes, GIFs, Videos - Visual Content* |  |  |  |  |

## Evidence for Online Civic Intervention Strategies

### I. Defuse & Distract

We identified five different methods of defusing and distracting attention away from TOC (Table 3). This set of responses defuses the tension and toxicity of a comment and distracts the speaker and/or the audience by using humor, mocking or ridiculing toxic content, exposing bot-like or for-profit behaviors, encouraging help-seeking and distracting with calming forms of content.

**Table 3:** Five ODE methods of using defusing and distracting

| Type of Response | Example |
|---|---|
| 1a. Using humor | Ah yes, it must be nice to have such a Hollywood view of the world. |
| 1b. Mocking or ridiculing | Did you not have your coffee yet today? Kindly take care of yourself |
| 1c. Exposing the bot-like behavior and/or profit motives of the person posting | Huh. Ya sound like a bot |
| 1d. Encourage help-seeking behaviors | Hey - there are people who can help you manage your anger. Find some help. |

| 1e. Distract with breathing or other calming content | This is a tense conversation. This video might help us take a deep breath and express issues without insults. |
|---|---|

Groups like #IAmHere and the Dangerous Speech Project identify humor as a TOC response strategy (1a). One study found that in online gaming settings, humor tends to ease hostility, offer support to other online speakers, and encourage social cohesion as a community-building “cushioning glue” that connects, seals, and buffers different gears of computer-mediated interaction (Marone, 2015). But other studies set in other contexts found humor unhelpful.

Through interviews with 12 specialists and a brief online experiment, researchers evaluated two one-minute videos responding to TOC. One presented a serious alternative ODE narrative and another used humor. They found that humor was not more effective in reducing harmful attitudes or promoting prosocial online behavior. In some cases, humor increased participants' tendency to react swiftly to hate content, raising concerns about its potential to amplify polarizing dynamics. Researchers caution that humor can offer benefits such as easing tension or prompting reflection, but it can also have negative effects by reinforcing stereotypes, alienating targets, or resonating with those who already agree. The results suggest that ODE narratives favoring long-term cultural change are more impactful than those using sarcastic humor (Choquette et al., 2024).

A case study of an organized “troll army” of online hate speech in Germany, Reconquista Germanica, and the counterspeech initiative Reconquista Internet explored the extent of different strategies in online media practices, with a specific focus on the interplay between users who spread hate and counterspeakers in the comments sections. Findings indicate that fact-building, as one of counter-building strategies, not only has limited effectiveness, but also generates opposing reactions and can even fuel tensions. However, this study found that using gentle humor helps lay down bridges between communities, lowers stress and social tensions, and sparks reflection, while a type of humor that humiliates the person can have a negative effect, as it makes users who post hate comments more resistant to change (Keller & Askanius, 2020).

Research sometimes finds that counterspeakers use ridicule (1b) as a default strategy to respond to TOC. Yet despite the widespread appeal and use of ridicule, or shaming of TOC speakers (1b), like humor, this strategy also has mixed results and may be context dependent. One research team used a combination of human annotators, language models, and machine learning to analyze TOC in 130,000 Twitter discussions over four years in individual tweets, discussion threads, and daily conversations. They found that sarcasm, irony, and cynicism strategies reduce extreme speech over time and lead to fewer politically extreme speakers in following comments. However, this can cause short-term hate increases in some groups(Lasser et al., 2023).

Exposing the bot-like behavior of someone posting TOC or exposing the profit motives of that person is another strategy (1c). Civic Health Project's Normsy.ai bot uses this strategy (Normsy.ai, 2025). One study found that human accounts were more likely to post content related to health and public welfare during the COVID-19 pandemic while *high bot score accounts* were more often promoting political conspiracies and divisive hashtags (Ferrara 2020 ), Another study found that flagging bot accounts on Twitter was an effective ODE strategy in reducing public acceptance of them (Lanius et al., 2021).

Promoting help-seeking behaviors (1d) is another response strategy to defuse or distract from TOC. A large body of research with individuals attracted to violent extremism (who might post online hate) found that they suffer loneliness and struggle with mental health (RAN, 2020). Encouraging a perpetrator of TOC to seek out mental or emotional support is a strategy emerging from counternarrative work of groups preventing violent extremism. Moonshot CVE, an NGO focused on preventing violent extremism, collaborated with Google to redirect users searching for violent content. When users entered specific search terms, they were shown options including mental health services, information countering hate, and nonviolent alternatives. Individuals viewing the redirect options were 48% more likely to click on an advertisement for emotional support than a control group (Uchill, 2018).

Finally, calming or breathing prompts (1e) might have an impact on TOC. Digital wellness programs suggest breathing exercises and calming content to address TOC (VODA, 2025). Stress inhibits the prefrontal cortex responsible for problem solving. Programs focused on breathing-based resilience can enhance psychological resilience, reduce stress, and diminish impulsive actions like writing TOC (Seppäl, et al., 2020). Slow breathing GIFs or calming digital content could help perpetrators find better ways to air grievances, enable byscrollers to recognize agency for ODE, and encourage victims to use breathing as coping. More research is needed to determine if posting GIFs promoting breathing exercises would have these effects.

## II. Engaging the Speaker's Perspective

A second category of responses aim to support public discourse norms by directly addressing the person who posted TOC to change their tone and reduce the level of toxicity (Table 4). Groups like Civic Health Project's Normsy.ai (2025) chatbot and Build Up use these strategies in their ODE programs (Build Up 2018).

**Table 4:** Five ODE methods of engaging the speaker's perspective

| Type of Response | Example |
|---|---|
| 2a. Reframe to express grievance in a less harmful way | It sounds like you are very worried about the state of the world. |
| 2b. Show comprehension with paraphrasing | If I understand you correctly, you are feeling frustrated about how the economy is impacting your life. |
| 2c. Show care and empathy to the speaker | I'm concerned for you and wonder if you might be able to keep your own dignity in how you decide to post |
| 2d. Ask questions, perhaps to provoke reflection | What experiences in your life led you to express yourself in this way? |
| 2e. Show support for certain parts of a post | I agree with you that this is an important issue. But let's not get so nasty. |

A substantial body of research in psychology and sociology documents the tendency for individuals to form "in-groups," to which they belong and feel positively inclined, and "out-groups," which they perceive more negatively (Tajfel & Turner, 1985). The preference for an in-group and hostility towards an out-group contribute to the propagation of online hate speech. In the study on moderating hate and extreme views on Twitter cited in the last section, researchers found that any reference to in-groups and out-groups exacerbates hate speech in online environments. Instead, sharing simple opinions, even without facts, can help counter harmful speech if they do not include insults or "othering" communication forms that are so common with ingroup/outgroup dynamics (Lasser et al., 2023). This category of ODE uses a variety of forms of perspective-taking.

A "reframing" response strategy (2a) involves taking a toxic comment and making it less toxic while keeping the core grievance or message. A set of researchers developed a "hate speech normalization" dataset called NACL (Neural hAte speeCh normaLizer), that first measures the hate intensity of a post, then identifies the specific hateful elements within it, and then reframes those elements to reduce the hate intensity while preserving the core meaning. A variety of metrics found that the NACL nudged perpetrators to reduce toxicity across platforms such as Reddit, Facebook, and Gab (Masud et al., 2022).

A paraphrasing strategy (2b) operates under the assumption that helping a perpetrator of TOC *feel heard* will encourage the person posting to communicate in a more productive way. In offline communication, decades of research document the significant impact that feeling heard has on a speaker (Livingstone et al., 2020). Feeling heard can enable an individual to move away from instinctual emotional reactions like anger or sadness and enable more acceptance of other viewpoints (Itzchakov & Weinstein, 2021).

Another response strategy is showing care and empathy for the TOC speaker (2c). A field experiment showed that LLM-generated recommendations for responding to online comments can improve political conversations by helping people express openness to their political opponents' views. The LLM offered suggestions for responding to "divisive online political conversations" based on three techniques: 1) listening and restatement to show comprehension; 2) validation to recognize the issue without necessarily agreement, and 3) using more polite language. LLMs were able to increase people's perception of feeling heard, and showed a higher willingness to understand opposing viewpoints in general, indicating a possible ripple effect (Argyle et al., 2023).

Researchers explored countering online hate speech through personalized counter-narratives (CNs) rather than generic responses. The authors found that integrating author profile information—age and gender—into language models like GPT-2 and GPT-3.5 produces messages that resonate better with hate speech authors. The findings highlight the promise of personalized automated ODE (Doganc & Markov, 2023).

In another study, researchers involved ten former violent extremists as "intervention providers" to write personal messages to individuals at risk of carrying out violence. The intervention providers employed various strategies, and researchers assessed the response rates as indicators of impact. Offers of assistance (2c) evoked responses 92% of the time, In comparison, 78% responded to personal stories, 42% to "ideological challenges," 36% to a "personal question" (2d), and 15% to "highlight the consequences of negative actions." This research suggests that some forms of engaging TOC.speakers are more effective than others. (Frenett & Dow, 2016, 17).

Even though one of the earliest studies of counterspeech found that asking questions (2d) was one of the most popular ODE response strategies (Bartlett & Krasodomski-Jones, 2015), few researchers test this strategy. We found no research on showing support for only parts of a post (2e), though our interviews found both the Civic Health Project and Smart Politics use these strategies.

## III. Identifying Shared Values

This set of responses aims to support public discourse norms by positive civility modeling, helping participants build on shared values of the need for safety, dignity, and identify where there is common ground on issues or values (Table 5). The first two strategies stem from the Smart Politics described earlier in this paper (Tamerius, 2019). If strategies such as establishing safety (3a) and setting dignity boundaries (3b) do not have impact, practitioners in improving public discourse norms end the conversation (Tamerius 2020). Conflict resolution practitioners emphasize the role of identifying shared values or ideas (3c). Researchers have proposed the concept of moral reframing (3d) as a specific approach to finding common ground. Similarly, religious actors such as Pope Francis advocate for responding to TOC with religious values (3e).

**Table 5:** Five ODE methods of identifying shared values

| Type of Response | Example |
|---|---|
| 3a. Establish safety | I want to understand you. So I won't attack or call you names. |
| 3b. Set dignity boundaries | If everyone here can respect the dignity of others, we might have a more productive conversation. |

| 3c. Identify shared values or ideas | I agree that the issue you named is a problem. Let’s think together how to solve it in a respectful way. |
|---|---|
| 3d. Use moral reframing | I agree with you that freedom and liberty are important, and this is why everyone needs to have the same rights. |
| 3e. Use religious values | In my religion, we think it is important to treat others the way we would want them to treat us. |

According to social cognitive theory, people acquire behavior through observational learning (Bandura, 2009). When people encounter civil behavior, they tend to emulate it, elevating conversation quality. Neuroscience research shows that humans possess mirror neuron systems; neural networks that fire both when we perform an action and when we observe someone else performing it (Acharya & Shukla, 2012). Neural mirroring underlies empathy, helping people sense others' emotions and intentions by "mirroring" them internally. Neural mirroring suggests that responses to TOC, whether empathetic or hostile, can directly shape how others feel and reply. When exposed to civil discussions, individuals are more likely to adopt civil discourse in their comments and show higher willingness to participate in future discussions.

An analysis of 15,000 Twitter and YouTube comments) identified that most counterspeech in the wild is argumentative using reasoning for facts, and contains hostile tones. While these posts garner more comments than positive counterspeech identifying shared values, the tone of the comments was more often argumentative. This suggests that typical metrics indicating ODE impacts should not rely on impressions or number of comments to indicate positive impact on public discourse. The authors conclude that only counter-speech which acknowledged grievances or displayed positive emotions resulted in positive dialogues, which they view as the ultimate aim of counterspeech (Baider, 2023).

A small study using college students as participants found exposure to civil discourse in online comments fostered more civil responses in online discussions, demonstrating that civility can be "contagious" (Han & Brazeal, 2015). The authors' later work, again with college students, confirmed that participants exposed to civil comments, defined as disagreeing with another’s comments while maintaining a respectful tone, were more likely to use civil discourse and stay on topic (Han et al., 2018).

Other researchers draw on research that people try to save face and expect politeness both in online and in-person communication. They find that any type of disagreement can result in more discomfort from others in participating in the conversation who perceive it to be impolite or threatening to a person's identity. Uncivil comments were more likely to prompt even more uncivil comments (Masullo Chen & Lu, 2017). This study might indicate that highlighting areas of common ground or shared values may be more effective than trying to disagree in a civil manner.

Several research studies have found that framing political arguments in terms of an audience’s own moral values can increase their persuasiveness. Rather than relying on data-heavy reasoning, effective moral reframing works by aligning messages with the values that people already hold, which improves processing fluency and affirms their moral identity (Feinberg and Willer, 2019). This approach builds directly on Jonathan Haidt’s moral foundations theory, which argues that people’s political judgments are rooted in distinct moral intuitions such as care, fairness, loyalty, authority, and sanctity (Haidt, 2013). Researchers conducted an online experiment conducted in South Korea found that comments reflecting core moral foundations were more persuasive than other types of comments (Suh & Kim, 2024).

Finally, religious actors, influencers, youth, and educators are beginning to promote *religious teachings, values, and language* (3e) as a response strategy to TOC. By drawing on shared moral principles such as dignity, compassion, justice, or the sanctity of human life, these responses reframe toxic discourse in ways that resonate with both insiders (those who share the faith) and outsiders. Some research suggests in some settings this can both *protect targeted youth* and *counter extremist misuses of religion* (Litvak et al., 2024).

## IV. Upstanding for Victims

This category of responses aims to support public discourse norms by speaking up for victims targeted by toxic posts (Table 6). These responses show empathy and solidarity to victims, invite others to do so, expose toxic posts, and warn of the consequences.

**Table 6:** Five ODE methods of upstanding for victims

| Type of Response | Example |
|---|---|
| 4a. Show empathy or solidarity with victims | This post is harmful to people I care about. |
| 4b. Invite others to upstand | I welcome others to join me in saying this is comment is harmful. |
| 4c. Denounce and expose TOC | The tone of this post is toxic. Please give others the dignity you would want to experience. |
| 4d. Warn of consequences | This type of post can lead to violence. Be careful. |
| 4e. Chime in on another's norm-setting | I add my support to this post. |

One study found that ODE that supports victims with simple messages of solidarity (4a) such as "I've got your back" could mitigate perceived harms and feelings of social isolation. This was especially salient when the counter-speaker is a member of the majority group  that shares the same traits as the perpetrator of the hate speech (Van Houtven et al., 2024).

A longitudinal analysis of over 130,000 Twitter "reply-tree" conversations in German political discourse from 2015 to 2018 between anti-immigrant group Reconquista Germanica (RG) and prosocial group Reconquista Internet (RI) found that counterspeech can reduce hate speech, but its effectiveness depends on bystander dynamics of inviting others to upstand (4b) and chiming in on norm setting (4e). Before organized efforts emerged, individual counter speakers dampened hate temporarily, yet their impact faded under sustained toxic content. When the RI movement mobilized, counterspeech became more visible and supportive, encouraging passive bystanders to join in and reinforcing civility norms. Importantly, byscrollers were not directly recruited; rather, RI's coordinated activity created a sense of safety that made engagement more likely, showing that organized collective counterspeech can shift online discourse more effectively than individual responses.(Garland et al., 2022).

In another experiment involving 1,350 English-speaking Twitter users who had posted xenophobic or racist content, researchers tested three strategies: empathy toward the victims of TOC, warning of consequences, and humor, against a control group that received no response. The results showed that empathy-based counterspeech was the only strategy to consistently reduce hateful behavior. Specifically, empathy messages prompted an increased deletion of the original hateful tweet and a reduction in future xenophobic content over four weeks. The other strategies of using humor and warnings of consequences showed no significant impact. These findings provide the first experimental evidence that empathy, rather than cynicism or fear, is a uniquely effective tool for mitigating online hate speech in real-world conditions (Hangartener et al., 2021).

In a study on Reddit, researchers tested three approaches to reduce hostile comments using an AI bot. The first was *descriptive norm induction*, reminding people what most others do, for example, saying avoiding insults. The second was *prescriptive norm induction*, telling people what they should do by asking them to be respectful. The third was *empathy induction*, encouraging people to consider how their words might hurt others. All three methods reduced aggressive comments, showing that simple reminders can make online conversations less toxic.(Seehausen et al., 2012).

One study found that ODE that addresses stereotypes against Chinese people and transgender people by standing up for these groups can increase polarization, potentially endangering social cohesion. Conflict between attackers and supporters of a minority group might amplify an us-vs-them narrative with left-wing and right-wing affiliations (Slothuus & de Vreese, 2010). The authors observe that those holding negative stereotypes toward attacked groups might experience a "boomerang effect" becoming more negative about these groups (Hart & Nisbet, 2012). ODE strategies aimed to increase byscrollers advocating for victims might unfortunately suffer from more rather than less toxic speech (Schäfer et al., 2023).

An experiment with individuals posting or searching for extremist content showed signs of behavioral impact redirecting individuals away from toxic content (Saltman et al., 2023). An experiment explored whether automated ODE criticizing TOC via bots could reduce racist harassment on Twitter. Automated accounts that varied by identity (white man vs. Black man) and follower count responded to posts with anti-Black slurs. High-follower white male bots significantly reduced their use of racist slurs over two months, showing that peer-group sanctions, especially from influential in-group sources, can curb hateful behavior online (Munger, 2016).

Researchers deployed an AI bot on Reddit to test three intervention types aimed at reducing verbal aggression. The bot detected users' aggressive comments and responded automatically with messages that highlighted community norms, appealed to civility ideals, or offered empathic understanding. All interventions significantly reduced verbal aggression compared to controls, with norm-based messages having the strongest effect, while empathy messages produced a smaller, but meaningful decline. The study shows that automated, psychologically-informed counterspeech can effectively reduce toxic language online, offering a scalable supplement to traditional moderation.(Bilewiz et al., 2021).

Several studies indicated that while byscrollers are able to identify toxic speech, many are not able to develop responses to it, suggesting the need for more training (Castellvi et al., 2022)

### V. Information & Fact-Building

The final category of response strategies supports public discourse norms aimed at understanding information sources and different views (Table 7). These responses show different information and erode the speaker's confidence in their views, including showing inconsistencies in their logic or that stereotypes reflect universal traits.

**Table 7:** Five ODE methods of information and fact building

| Type of Response | Example |
|---|---|
| 5a. Offer news links that show different perspectives | Here is another news source that seems to show different information. |
| 5b. Offer science or legal evidence | The scientific/legal evidence for you point is weak. Take a look at this report. |
| 5c. Erode the speaker's intellectual framework | What is your source of information? Critical thinking is important, and this post doesn't make sense. |

| 5d. Show inconsistencies in logic | Your posts seem to hold a double standard and mixed message. Try checking your sources, just like you ask others to check theirs. |
|---|---|
| 5e. Show universal traits | This stereotype doesn't make sense since these traits are found in every group of people. |

A variety of research reports offer strategies for responding to TOC that uses false or deceptive information (Lewandowsky et al., 2020). In counterspeech research, some researchers group all fact-based response strategies together, so it is difficult to assess if one of the response strategies is more effective than another. Overall, the research affirms the value of this category of responses.

Some ODE research offers news from different perspectives (5a) with tools to help people understand how different journalists are covering the same issue or story. Research found that showing people their blindspots can, under certain conditions, may shift perspectives (Bram, 2024). For example, AllSides presents headlines from news sources rated as left, center, and right and 14 variations of bias, allowing readers to compare how different outlets frame the same issue. AllSides uses crowd-sourced and editorial reviews to determine its media bias ratings (Mastrine et al., 2019). Ground News is a "news comparison tool" that presents stories side-by-side from multiple outlets. It allows users to filter results by political bias (left, right, or center), location, and whether a source is based in the U.S. or abroad. A feature called "Blindspot" shows stories that receive little or no coverage from one side of the political spectrum.

A study on human-generated factual and sarcastic replies from journalists to uncivil comments on news stories found that factual responses enhanced readers' perception of a productive conversation and increased their willingness to participate in the conversation. Factual, polite replies from journalists to uncivil comments were effective at setting norms for respectful dialogue, while sarcastic responses backfired by undermining credibility and discouraging engagement. Conversely, sarcastic responses, while potentially entertaining, undermined the credibility of the news outlet and the perceived quality of its journalism, ultimately reducing readers' willingness to engage (Ziegele & Jost, 2016).

A study on AI automated counter-narratives found that fact-based responses using document retrieval and summarization are more persuasive, trustworthy, and coherent than generic or emotional replies. Unlike Ziegele and Jost's human moderation study, Wilk et al. show that automation can scale fact-based responses when factual grounding is ensured. These studies suggest that across human and AI-mediated contexts, fact-based counterspeech outperforms sarcastic or emotional strategies (Wilk et al., 2025).

A study asked annotators to evaluate counterspeech quality from expert-written counterspeech (from the CONAN dataset) and user-generated counterspeech from Twitter across six dimensions: clarity, evidence, rebuttal, fairness, emotional appeal, and audience adaptation. Expert-generated responses provided stronger factual backing, making them more persuasive, while user-generated responses relied more on emotion but lacked evidence.(Damo et al., 2025).

Extensive research on TOC related to vaccines and climate suggests that simply stating the scientific agreement or citing scientific evidence (5b) does not effectively change the views of those who are skeptical (Farrell et al., 2019). However prebunking or informing people about false information *before* they encounter it by inoculating them with information about counterarguments, a form of ODE, does show promise (van der Linden et al., 2017)

ODE strategies that aim to erode the speaker's intellectual framework (5c) or show inconsistencies (5d) are less researched. In a 2023 study on Twitter, researchers discovered that sharing straightforward opinions, even if not backed by evidence but free from insults, is linked to the least amount of hate in ensuing discussions. Additionally,

ODE that includes a strong emotional tone, whether negative like anger or fear, or positive like enthusiasm and pride, tends to result in poorer discourse quality (Lasser et al., 2023), perhaps a signal that a more neutral toned, fact-based ODE might be more effective.

The final strategy addresses TOC that expresses prejudices against an entire group of people, such as migrants, Muslims, Jews, or women, with assertions that traits are "universal" (5e). One study judging automated ODE found this strategy effective. Researchers asked male and female participants to assess AI-generated gender-based responses to gender stereotypes. Participants compare eleven automated ODE response strategies to assess offensiveness, plausibility, and effectiveness. Participants found that automated responses using counter-facts and "broadening universals" (i.e., stating anyone can have a trait regardless of group membership) emerged as the most robust approaches, while humor, perspective-taking, counter-examples, and empathy were perceived as less effective strategies. Results suggest that AI-attributed responses grounded in facts and universality are strongest for countering biased statements, even when automatically generated.(Nejadgholi et al., 2024).

## Limitations

This study faces several limitations. First, there are search and selection biases. Our keyword-based literature search may have produced a narrowed sample, excluding studies using different terms to describe response strategies to toxic online content (TOC). The scope's intentional limitation to TOC may have excluded important adjacent work in social work or psychology that could provide insights into online discourse engagement. Additionally, there is a language barrier. Our English search terms and reliance on English publications means important studies in other languages may have been overlooked. We included Portuguese, Urdu, and Spanish language publications, but researchers did not conduct thorough literature review in these languages.

Second, subjectivity and time constraints are inherent in this literature review. Decisions about which papers to include, how to code them, and how to classify response strategies involved researcher judgment and interpretation. The review captures evidence only up to data collection; relevant studies published after our cutoff could further expand the taxonomy.

Third, the literature reviewed itself uses different research methodologies, making it impossible to compare outcomes. Some papers measure outcomes using weak proxies like "impressions" or engagement which do not capture attitudinal or behavioral change. Few studies employ rigorous experimental or longitudinal methods. Different researchers use inconsistent definitions, taxonomies, and metrics, making it difficult to compare findings across studies. Consequently, our taxonomy and evaluation may unintentionally reflect gaps and biases within the existing literature.

## Conclusion

This paper offers a comprehensive taxonomy of online discourse engagement (ODE) in response to toxic online content (TOC). By synthesizing scholarship across disciplines, we have demonstrated that ODE encompasses a wide spectrum of strategies—ranging from humor and distraction to empathy, fact-based rebuttals, and moral reframing. Our taxonomy clarifies definitions, distinguishes goals, and highlights the diversity of strategies employed by individuals, NGOs, governments, and AI systems. This framework provides a foundation for more systematic evaluation and comparative analysis, enabling researchers and practitioners to speak with greater precision about what kinds of responses are being studied and implemented.

The evidence base we reviewed suggests that no single strategy is universally effective, and there is scant evidence for any of the 25 strategies identified. In general, fact-based and empathy-based counterspeech toward both speakers and victims of TOC shows strong promise in reducing harmful speech. In general, strategies rooted in humor or sarcasm may backfire if perceived as insincere or offensive. Similarly, fact-based rebuttals can enhance perceptions of deliberation but risk escalating hostility if framed antagonistically.

Research outcomes vary across cultural contexts, platform dynamics, and the identity of both speakers and responders. More research is necessary. Our future research plans to evaluate these different strategies in an experimental context and survey practitioners to better understand which strategies they believe are most effective.

Collective interventions, such as #iamhere or Reconquista Internet, appear to magnify impact by shifting norms among byscrollers (online bystanders), illustrating the importance of scale and coordination. These findings underscore the need to align ODE strategies with clearly defined goals, whether supporting victims, dissuading bystanders, or engaging perpetrators.

Further research is also needed to identify metrics that measure success through changes not just in "impressions" or engagement on social media but also with new research methodologies that can measure shifts in attitudes, beliefs, and behaviors.

New LLM-boosted ODE such as Civic Health Project's Normsy AI chatbot (Normsy.ai, 2025) and Plurality Institute's Bridging Bot (Plurality Institute, 2025) offer new opportunities to collect data and then compare and contrast which ODE strategies are most effective. As researchers advance fine-tuning of LLMs to perform ODE, it is urgent that the emerging field of AI-enhanced response to TOC do the necessary field building work of developing shared terminology, taxonomies and definitions. This paper contributes to this task.

Advancing ODE will require stronger methodological rigor, cross-disciplinary collaboration, and participatory research integrating practitioner insights with empirical testing. Comparative studies across platforms and cultures are essential to determine when strategies succeed or fail. Policymakers and funders must recognize ODE as an ecosystem where multiple approaches interact, and supporting diverse strategies will be more effective than privileging single tactics. ODE aims to cultivate prosocial discourse norms that affirm human dignity, foster deliberation, and strengthen democratic values for healthy societies.

## Acknowledgments:

The authors would like to thank graduate students from Keough School of Global Affairs at the University of Notre Dame for their help in collecting and organizing the literature including Julie Hawke, Grace Connor, Miriam Bethencourt, Nik Swift, Maria Kobayashi Rossi, Rida Ejaz, and Furqan Mohammed; Paul Brenner from Center for Research Computing at the University of Notre Dame for his support and continuous discussions throughout the project.